\documentstyle[amssymb,floats,aps,prb,epsf]{revtex}
\epsfclipon
\begin{document}
\twocolumn[\hsize\textwidth\columnwidth\hsize\csname
@twocolumnfalse\endcsname

\draft
\title{Superconductivity in doped two-leg ladder cuprates}

\author{Jihong Qin}

\address{Department of Physics, Beijing University of Science
and Technology, Beijing 100083, China}

\author{Feng Yuan}
\address{Department of Physics, Qingdao University, Qingdao
266071, China}

\author{Shiping Feng}

\address{Department of Physics, Beijing Normal University,
Beijing 100875, China}

\maketitle
\begin{abstract}
Within the $t$-$J$ ladder model, superconductivity with a modified
d-wave symmetry in doped two-leg ladder cuprates is investigated
based on the kinetic energy driven superconducting mechanism. It
is shown that the spin-liquid ground-state at the half-filling
evolves into the superconducting ground-state upon doping. In
analogy to the doping dependence of the superconducting transition
temperature in the planar cuprate superconductors, the
superconducting transition temperature in doped two-leg ladder
cuprates increases with increasing doping in the underdoped
regime, and reaches a maximum in the optimal doping, then
decreases in the overdoped regime.
\end{abstract}
\pacs{74.20.Mn,74.20.Rp,74.25.Dw}

]
\bigskip
\narrowtext


In recent years two-leg ladder cuprates have attracted great
interest since their ground state may be a spin liquid state with
a finite spin gap \cite{dagotto1,dagotto2,katano}. This spin
liquid state may play a crucial role in superconductivity of the
planar cuprate superconductors as emphasized by Anderson
\cite{anderson}. When carriers are doped into two-leg ladder
cuprates, a metal-insulator transition occurs
\cite{uehara,nagata,isobe}. Although the ambient pressure ladder
superconductivity was not oberved until now, superconductivity in
one of the doped two-leg ladder cuprate
Sr$_{14-x}$Ca$_{x}$Cu$_{24}$O$_{41}$ has been observed under high
pressure \cite{uehara,nagata,isobe}, which is the only known
superconducting (SC) copper oxide without a square lattice. It has
been shown that most important role of pressure for realizing
superconductivity is the doped hole redistribution between chains
and ladders \cite{uehara,nagata,isobe}, and then the number of
charge carriers on the ladders is increased \cite{ohta,kato}.
Moreover, the structure under high pressure remains the same as
the case in ambient pressure \cite{isobe}, and the spin background
in this SC phase does not drastically alter its spin gap
properties \cite{dagotto2}. The particular geometrical arrangement
of the Cu ions in two-leg ladder cuprates provides a playground
for the normal- and SC-state studies of low-dimensional strongly
correlated materials \cite{dagotto1,dagotto2,katano}. This follows
from the fact that all planar cuprate superconductors found up now
contain square CuO$_{2}$ planes \cite{kastner}, whereas doped
two-leg ladder cuprates consists of two-leg ladders of other Cu
ions and edge-sharing CuO$_{2}$ chains
\cite{dagotto1,dagotto2,katano}. By virtue of the nuclear magnetic
resonance and nuclear quadrupole resonance, particularly inelastic
neutron scattering measurements, it has been shown that there is a
region of parameter space and doping where doped two-leg ladder
cuprates in the normal state is an antiferromagnet with the
commensurate short-range order \cite{katano,magishi}. Moreover,
transport measurements on doped two-leg ladder cuprates in the
same region of parameter space and doping indicate that the
resistivity is linear with temperatures \cite{nagata}, one of the
hallmarks of the exotic normal state properties found in the
planar cuprate superconductors \cite{kastner}. These experimental
results have revealed some close analogies between the doped
planar cuprates and doped two-leg ladder cuprates. The
normal-state of doped two-leg ladder cuprates exhibits a number of
anomalous properties which is due to the charge-spin separation
(CSS), while the SC state may be characterized by the charge-spin
recombination.

Theoretically, there is a general consensus that the charge
carrier pair of doped two-leg ladder cuprates in the SC-state are
created on the ladders \cite{dagotto2}, i.e., superconductivity
develops mainly within the ladders, with a minor role played by
the interladder hopping amplitude. Within the $t$-$J$ ladder
model, many authors have shown that the charge carrier pair
correlation is very robust \cite{dagotto3}, clearly indicative of
a ground-state dominated by strong SC tendencies. Moreover, it has
been shown in the renormalized mean-field (MF) theory that
superconductivity should exist in the d-wave channel
\cite{sigrist}, which has been confirmed by variety of numerical
simulations \cite{riera}. Within the framework of the CSS
fermion-spin theory \cite{feng1}, we have developed a kinetic
energy driven SC mechanism \cite{feng2}, where the dressed holons
interact occurring directly through the kinetic energy by
exchanging the spin excitations, leading to a net attractive force
between the dressed holons, then the electron Cooper pairs
originating from the dressed holon pairing state are due to the
charge-spin recombination, and their condensation reveals the SC
ground-state. This SC-state is controlled by both SC gap function
and quasiparticle coherence, then the maximal SC transition
temperature occurs around the optimal doping, and decreases in
both underdoped and overdoped regimes \cite{feng3}. In particular,
this kinetic energy driven SC mechanism does not depend on the
fine details of a lattice structure, and the main ingredient was
identified into a charge carrier pairing mechanism involving the
internal spin degree of freedom \cite{feng2}. Therefore this SC
mechanism shows that the strong electron correlation favors
superconductivity \cite{feng2,feng3}. Since there is a remarkable
resemblance in the normal-state properties between the doped
planar cuprates and doped two-leg ladder cuprates as mentioned
above, and the strong electron correlation is common for both
these cuprate materials, then two systems may have similar
underlying SC mechanism, i.e., it is possible that
superconductivity in doped two-leg ladder cuprates is also driven
by the kinetic energy. In this paper, we discuss superconductivity
in doped two-leg ladder cuprates along with the kinetic energy
driven SC mechanism. We show that in analogy to the doping
dependence of the SC transition temperature in the planar cuprate
superconductors, the SC transition temperature in doped two-leg
ladder cuprates increases with increasing doping in the underdoped
regime, and reaches a maximum in the optimal doping, then
decreases in the overdoped regime.

The basic element of two-leg ladder materials is the two-leg
ladder, which is defined as two parallel chains of ions, with
bonds among them such that the interchain coupling is comparable
in strength to the couplings along the chains, while the coupling
between the two chains that participates in this structure is
through rungs \cite{dagotto1,dagotto2}. In this case, the $t$-$J$
ladder model on the two-leg ladder is expressed as,
\begin{eqnarray}
H&=&-t_{\parallel}\sum_{i\hat{\eta}a\sigma}C_{ia\sigma}^{\dagger}
C_{i+\hat{\eta}a\sigma}-t_{\perp}\sum_{i\sigma}
(C_{i1\sigma}^{\dagger}C_{i2\sigma}+{\rm H.c.}) \nonumber \\
&-&\mu\sum_{ia\sigma}C_{ia\sigma }^{\dagger}C_{ia\sigma }+
J_{\parallel}\sum_{i\hat{\eta}a}{\bf S}_{ia}\cdot {\bf
S}_{i+\hat{\eta}a}\nonumber \\
&+&J_{\perp}\sum_{i}{\bf S}_{i1}\cdot {\bf S}_{i2},
\end{eqnarray}
where $\hat{\eta}=\pm c_{0}\hat{x}$, $c_{0}$ is the lattice
constant of the two-leg ladder lattice, which is set as unity
hereafter, $i$ runs over all rungs, $\sigma(=\uparrow,\downarrow)$
and $a(=1,2)$ are spin and leg indices, respectively,
$C^{\dagger}_{ia\sigma}$ ($C_{ia\sigma}$) are the electron
creation (annihilation) operators, ${\bf S}_{ia}=C^{\dagger}_{ia}
\vec{\sigma}C_{ia}/2$ are the spin operators with ${\vec\sigma}=
(\sigma_{x},\sigma_{y},\sigma_{z})$ as the Pauli matrices, and
$\mu$ is the chemical potential. This $t$-$J$ ladder Hamiltonian
(1) is subject to an important on-site local constraint
$\sum_{\sigma}C_{ia\sigma}^{\dagger}C_{ia\sigma}\leq 1$ to avoid
the double occupancy. In the materials of interest
\cite{uehara,nagata,isobe}, the exchange coupling $J_{\parallel}$
along the legs is close to the exchange coupling $J_{\perp}$
across a rung, and the same is true of the hopping $t_{\parallel}$
along the legs and the rung hopping strength $t_{\perp}$.
Therefore, in the following discussions, we will work with the
isotropic system $J_{\perp}=J_{\parallel}=J$, $t_{\perp}=
t_{\parallel}=t$. On the other hand, the single occupancy local
constraint in the $t$-$J$ ladder Hamiltonian (1) can be treated
properly in analytical calculations within the CSS fermion-spin
theory \cite{feng1}, $C_{ia\uparrow}=h^{\dagger}_{ia\uparrow}
S^{-}_{ia}$, $C_{ia\downarrow}=h^{\dagger}_{ia\downarrow}
S^{+}_{ia}$, where the spinful fermion operator $h_{ia\sigma}=
e^{-i\Phi_{i\sigma}}h_{ia}$ describes the charge degree of freedom
together with some effects of the spin configuration
rearrangements due to the presence of the doped hole itself
(dressed holon), while the spin operator $S_{ia}$ describes the
spin degree of freedom (spin), then the electron local constraint
for single occupancy, $\sum_{\sigma} C^{\dagger}_{ia\sigma}
C_{ia\sigma}=S^{+}_{ia}h_{ia\uparrow} h^{\dagger}_{ia\uparrow}
S^{-}_{ia} +S^{-}_{ia}h_{ia\downarrow} h^{\dagger}_{ia\downarrow}
S^{+}_{ia}=h_{ia} h^{\dagger}_{ia} (S^{+}_{ia}S^{-}_{ia}+
S^{-}_{ia}S^{+}_{ia})=1-h^{\dagger}_{ia} h_{ia}\leq 1$, is
satisfied in analytical calculations, and the double spinful
fermion occupancies $h^{\dagger}_{ia\sigma}
h^{\dagger}_{ia-\sigma}=e^{i\Phi_{i\sigma}} h^{\dagger}_{ia}
h^{\dagger}_{ia}e^{i\Phi_{i-\sigma}}=0$ and $h_{ia\sigma}
h_{ia-\sigma}=e^{-i\Phi_{i\sigma}}h_{ia}h_{ia}
e^{-i\Phi_{i-\sigma}}=0$, are ruled out automatically. Since these
dressed holons and spins are gauge invariant, and then in this
sense, they are real and can be interpreted as the physical
excitations \cite{laughlin}. Although in common sense
$h_{ia\sigma}$ is not a real spinful fermion, it behaves like a
spinful fermion. In this CSS fermion-spin representation, the
low-energy behavior of the $t$-$J$ ladder Hamiltonian (1) can be
expressed as,
\begin{eqnarray}
H&=&t\sum_{i\hat{\eta}a}(h^{\dagger}_{i+\hat{\eta}a\uparrow}
h_{ia\uparrow}S^{+}_{ia}S^{-}_{i+\hat{\eta}a}+
h^{\dagger}_{i+\hat{\eta}a\downarrow}h_{ia\downarrow}S^{-}_{ia}
S^{+}_{i+\hat{\eta}a}) \nonumber \\
&+&t\sum_{i}(h^{\dagger}_{i2\uparrow}h_{i1\uparrow}S^{+}_{i1}
S^{-}_{i2}+ h^{\dagger}_{i1\uparrow}h_{i2\uparrow}S^{+}_{i2}
S^{-}_{i1}\nonumber\\
&+& h^{\dagger}_{i2\downarrow}h_{i1\downarrow}S^{-}_{i1}
S^{+}_{i2}+h^{\dagger}_{i1\downarrow}h_{i2\downarrow}S^{-}_{i2}
S^{+}_{i1}) \nonumber \\
&+&\mu\sum_{ia\sigma}h^{\dagger}_{ia\sigma}h_{ia\sigma} +{J_{\rm
eff}}\sum_{i\hat{\eta}a}{\bf S}_{ia}\cdot {\bf S}_{i+\hat{\eta}a}
\nonumber \\
&+& {J_{\rm eff}}\sum_{i}{\bf S}_{i1}\cdot {\bf S}_{i2},
\end{eqnarray}
with $J_{\rm eff}=J(1-\delta)^{2}$, and $\delta=\langle
h^{\dagger}_{ia\sigma}h_{ia\sigma}\rangle=\langle h^{\dagger}_{ia}
h_{ia}\rangle$ is the hole doping concentration. In this CSS
fermion-spin representation, the kinetic terms have been expressed
as the dressed holon-spin interactions, which reflect that even
the kinetic energy terms in the $t$-$J$ ladder Hamiltonian have
the strong Coulombic contribution due to the restriction of no
doubly occupancy of a given site. As in the planar cuprate
superconductors \cite{tsuei}, the SC state in doped two-leg ladder
cuprates is also characterized by electron Cooper pairs
\cite{uehara,nagata,isobe}, forming SC quasiparticles. Because
there are two coupled $t$-$J$ chains in the two-leg ladder
cuprates, therefore the order parameters for the electron Cooper
pair is a matrix $\Delta=\Delta_{L}+\sigma_{x}\Delta_{T}$, with
the longitudinal and transverse order parameters are defined as,
\begin{mathletters}
\begin{eqnarray}
\Delta_{L}&=&\langle C^{\dagger}_{ia\uparrow}
C^{\dagger}_{ja\downarrow}-C^{\dagger}_{ia\downarrow}
C^{\dagger}_{ja\uparrow}\rangle\nonumber \\
&=&\langle h_{ia\uparrow} h_{ja\downarrow}S^{+}_{ia}
S^{-}_{ja}-h_{ia\downarrow} h_{ja\uparrow}S^{-}_{ia}
S^{+}_{ja}\rangle \nonumber \\
&=&-\langle S^{+}_{ia}S^{-}_{ja}\rangle\Delta_{hL},\\
\Delta_{T}&=&\langle C^{\dagger}_{i1\uparrow}
C^{\dagger}_{i2\downarrow}- C^{\dagger}_{i1\downarrow}
C^{\dagger}_{i2\uparrow}\rangle \nonumber\\
&=&\langle h_{i1\uparrow}
h_{i2\downarrow}S^{+}_{i1}S^{-}_{i2}-h_{i1\downarrow}
h_{i2\uparrow}S^{-}_{i1}S^{+}_{i2}\rangle \nonumber \\
&=&-\langle S^{+}_{i1} S^{-}_{i2}\rangle\Delta_{hT},
\end{eqnarray}
\end{mathletters}
respectively, where the longitudinal and transverse dressed holon
pairing order parameters are expressed as,
\begin{mathletters}
\begin{eqnarray}
\Delta_{hL}=\langle h_{ja\downarrow}h_{ia\uparrow}-h_{ja\uparrow}
h_{ia\downarrow}\rangle, \\
\Delta_{hT}=\langle h_{i2\downarrow}h_{i1\uparrow}-h_{i2\uparrow}
h_{i1\downarrow}\rangle.
\end{eqnarray}
\end{mathletters}
In this case, the physical properties of doped two-leg ladder
cuprates in the SC state are essentially determined by the dressed
holon pairing state, i.e., the SC order parameters are determined
by the dressed holon pairing amplitude, and are proportional to
the number of doped holes, and not to the number of electrons.

For discussions of superconductivity in the doped two-leg ladder
cuprates, we now introduce the dressed holon normal and anomalous
Green's functions and spin Green's functions as, $g({\bf k},
\omega) =g_{L}({\bf k},\omega)+\sigma_{x} g_{T}({\bf k},\omega)$,
$\Im^{\dagger}({\bf k},\omega)= \Im^{\dagger}_{L}({\bf k},\omega)
+\sigma_{x}\Im^{\dagger}_{T} ({\bf k},\omega)$, $D({\bf k},\omega)
=D_{L}({\bf k},\omega)+ \sigma_{x}D_{T}({\bf k},\omega)$, and
$D_{z}({\bf k},\omega)= D_{zL}({\bf k},\omega)+\sigma_{x}
D_{zT}({\bf k},\omega)$, respectively, where the longitudinal and
transverse parts of these Green's functions are defined as,
\begin{mathletters}
\begin{eqnarray}
g_{L}(i-j,t-t')&=&\langle\langle h_{ia\sigma}(t);
h^{\dagger}_{ja\sigma}(t')\rangle\rangle ,\\
g_{T}(i-j,t-t')&=&\langle\langle h_{ia\sigma}(t);
h^{\dagger}_{ja'\sigma}(t')\rangle\rangle ,\\
\Im^{\dagger}_{L}(i-j,t-t')&=&\langle\langle
h^{\dagger}_{ia\uparrow}(t);h^{\dagger}_{ja\downarrow}(t')
\rangle\rangle ,\\
\Im^{\dagger}_{T}(i-j,t-t')&=&\langle\langle
h^{\dagger}_{ia'\uparrow}(t);h^{\dagger}_{ja\downarrow}(t')
\rangle\rangle ,\\
D_{L}(i-j,t-t')&=&\langle\langle S^{+}_{ia}(t);S^{-}_{ja}(t')
\rangle\rangle, \\
D_{T}(i-j,t-t')&=&\langle\langle S^{+}_{ia}(t);S^{-}_{ja'}(t')
\rangle\rangle, \\
D_{zL}(i-j,t-t')&=&\langle\langle S^{z}_{ia}(t);S^{z}_{ja}(t')
\rangle\rangle, \\
D_{zT}(i-j,t-t')&=&\langle\langle S^{z}_{ia}(t);S^{z}_{ja'}(t')
\rangle\rangle ,
\end{eqnarray}
\end{mathletters}
with $a'\neq a$. In the MF level, the spin system is an
anisotropic away from the half-filling \cite{qin}, therefore we
have defined the two spin Green's functions $D({\bf k},\omega)$
and $D_{z}({\bf k},\omega)$ to describe the spin propagations,
while the longitudinal and transverse parts of the MF dressed
holon normal Green's function and MF spin Green's functions have
been obtained as \cite{qin},
\begin{mathletters}
\begin{eqnarray}
g^{(0)}_{L}({\bf k},\omega)&=&{1\over 2}\sum_{\nu=1,2}
{1\over \omega-\xi_{\nu{\bf k}}}, \\
g^{(0)}_{T}({\bf k},\omega)&=&{1\over 2}\sum_{\nu=1,2}
(-1)^{\nu+1}{1\over \omega-\xi_{\nu{\bf k}}},\\
D^{(0)}_{L}({\bf k},\omega)&=&{1\over 2}\sum_{\nu=1,2}
{B_{\nu{\bf k}}\over \omega^{2}-\omega^{2}_{\nu{\bf k}}}, \\
D^{(0)}_{T}({\bf k},\omega)&=&{1\over 2}
\sum_{\nu=1,2}(-1)^{\nu+1}{B_{\nu{\bf k}}\over \omega^{2}-
\omega^{2}_{\nu{\bf k}}}, \\
D^{(0)}_{zL}({\bf k},\omega)&=&{1\over 2}\sum_{\nu=1,2}
{B_{z\nu{\bf k}}\over\omega^{2}-\omega^{2}_{z\nu{\bf k}}}, \\
D^{(0)}_{zT}({\bf k},\omega)&=&{1\over 2}
\sum_{\nu=1,2}(-1)^{\nu+1}{B_{z\nu{\bf k}}\over\omega^{2}-
\omega^{2}_{z\nu{\bf k}}},
\end{eqnarray}
\end{mathletters}
where $B_{\nu {\bf k}}=\lambda [A_{1}{\rm cos}k_{x}- A_{2}]-
J_{\rm eff}[\chi_{\perp}+2\chi^{z}_{\perp}(-1)^{\nu}]
[\epsilon_{\perp}+(-1)^{\nu}]$, $B_{z\nu {\bf k}}=-\lambda
\epsilon_{\parallel}\chi_{\parallel}(1-{\rm cos}k_{x})+J_{\rm eff}
\epsilon_{\perp}\chi_{\perp}[(-1)^{\nu+1}-1]$, $\lambda=4J_{\rm
eff}$, $A_{1}=2\epsilon_{\parallel}\chi^{z}_{\parallel}+
\chi_{\parallel}$, $A_{2}=\epsilon_{\parallel}\chi_{\parallel}+
2\chi^{z}_{\parallel}$, $\epsilon_{\parallel}= 1+2t
\phi_{\parallel}/J_{\rm eff}$, $\epsilon_{\perp}=1+ 4t
\phi_{\perp}/J_{\rm eff}$, the spin correlation functions
$\chi_{\parallel}=\langle S_{ai}^{+} S_{ai+\hat{\eta}}^{-}
\rangle$, $\chi^{z}_{\parallel}=\langle S_{ai}^{z}
S_{ai+\hat{\eta}}^{z}\rangle$, $\chi_{\perp}=\langle S^{+}_{1i}
S^{-}_{2i}\rangle$, $\chi^{z}_{\perp}=\langle S_{1i}^{z}
S_{2i}^{z}\rangle$, the dressed holon particle-hole order
parameters $\phi_{\parallel}=\langle h^{\dagger}_{ai\sigma}
h_{ai+\hat{\eta}\sigma}\rangle$, $\phi_{\perp}=\langle
h^{\dagger}_{1i\sigma}h_{2i\sigma}\rangle$, and the MF dressed
holon and spin excitation spectra,
\begin{mathletters}
\begin{eqnarray}
\xi_{\nu{\bf k}}&=&2t\chi_{\parallel}{\rm cos}k_{x}+\mu+
\chi_{\perp}t(-1)^{\nu+1}, \\
\omega^{2}_{\nu{\bf k}}&=&{1\over 2}\alpha\epsilon_{\parallel}
\lambda^{2}A_{1}{\rm cos}^{2}k_{x}+[X_{1}+X_{2}(-1)^{\nu+1}]
{\rm cos}k_{x}\nonumber \\
&+&X_{3}+X_{4}(-1)^{\nu+1}, \\
\omega^{2}_{z\nu{\bf k}}&=&\alpha\epsilon_{\parallel}\lambda^{2}
\chi_{\parallel}{\rm cos}^{2}k_{x}+ [Y_{1}+ Y_{2}(-1)^{\nu+1}]
{\rm cos}k_{x} \nonumber\\
&+& Y_{3}+Y_{4}(-1)^{\nu+1},
\end{eqnarray}
\end{mathletters}
where $X_{1}=-\epsilon_{\parallel}\lambda^{2}[(\alpha
A_{2}+2A_{4})/4+A_{3}]-\alpha\lambda J_{\rm eff}
[\epsilon_{\parallel}(C^{z}_{\perp}+\chi^{z}_{\perp})+
\epsilon_{\perp}(C_{\perp}+\epsilon_{\parallel}\chi_{\perp})/2]$,
$X_{2}=\alpha\lambda J_{\rm eff}[(\epsilon_{\perp}
\chi_{\parallel}+\epsilon_{\parallel}\chi_{\perp})/2+
\epsilon_{\parallel}\epsilon_{\perp}(\chi^{z}_{\perp}+
\chi^{z}_{\parallel})]$, $X_{3}=\lambda^{2}[A_{3}-\alpha
\epsilon_{\parallel}A_{1}/4+\epsilon^{2}_{\parallel}A_{4}/2]+
\alpha\lambda J_{\rm eff}[\epsilon_{\parallel}\epsilon_{\perp}
C_{\perp}+ 2C^{z}_{\perp}]+J^{2}_{\rm eff}
(\epsilon^{2}_{\perp}+1)/4$, $X_{4}=-\alpha\lambda J_{\rm eff}
[\epsilon_{\parallel}\epsilon_{\perp}\chi_{\parallel}/2+
\epsilon_{\perp}(\chi^{z}_{\parallel}+C^{z}_{\perp})+
\epsilon_{\parallel}C_{\perp}/2]-\epsilon_{\perp}J^{2}_{\rm eff}
/2$, $Y_{1}=- \epsilon_{\parallel}\lambda^{2}
[\epsilon_{\parallel}(\alpha C_{\parallel}+(1-\alpha)/4)+
\alpha\chi_{\parallel}/2] -\alpha\lambda J_{\rm eff}
\epsilon_{\perp}(\epsilon_{\parallel} C_{\perp}+\chi_{\perp})$,
$Y_{2}=\alpha\lambda J_{\rm eff} (\epsilon_{\parallel}
\chi_{\parallel}+\epsilon_{\perp} \chi_{\perp})$, $Y_{3}=
\epsilon_{\parallel}\lambda^{2} [\epsilon_{\parallel}(\alpha
C_{\parallel}+(1-\alpha)/4)- \alpha \chi_{\parallel}/2]+
2\alpha\lambda J_{\rm eff} \epsilon_{\parallel}\epsilon_{\perp}
C_{\perp}+ \epsilon^{2}_{\perp}J^{2}_{\rm eff}/2$,
$Y_{4}=-\alpha\lambda J_{\rm eff}\epsilon_{\parallel}
(\chi_{\parallel}+\epsilon_{\perp} C_{\perp})-
\epsilon^{2}_{\perp}J^{2}_{\rm eff}/2$, with $A_{3}=\alpha
C^{z}_{\parallel}+(1-\alpha)/8$, $A_{4}=\alpha
C_{\parallel}+(1-\alpha)/4$, and the spin correlation functions
$C_{\parallel}= \sum_{\hat{\eta}\hat{\eta'}}\langle
S_{ai+\hat{\eta}}^{+} S_{ai+\hat{\eta'}}^{-}\rangle/4$,
$C^{z}_{\parallel}= \sum_{\hat{\eta}\hat{\eta'}}\langle
S_{ai+\hat{\eta}}^{z} S_{ai+\hat{\eta'}}^{z}\rangle/4$,
$C_{\perp}=\sum_{\hat{\eta}} \langle S_{2i}^{+}
S_{1i+\hat{\eta}}^{-}\rangle/2$, and $C^{z}_{\perp}=
\sum_{\hat{\eta}}\langle S_{1i}^{z}S_{2i+
\hat{\eta}}^{z}\rangle/2$.

Within the Eliashberg's strong coupling theory \cite{eliashberg},
it has been shown that the dressed holon-spin interaction in the
doped planar cuprates can induce the dressed holon pairing state
(then the electron Cooper pairing state) by exchanging the spin
excitations in the higher power of the doping concentration
\cite{feng2}. Following their discussions \cite{feng2}, the
self-consistent equations that satisfied by the full dressed holon
normal and anomalous Green's functions in doped two-leg ladder
cuprates are expressed as \cite{eliashberg},
\begin{mathletters}
\begin{eqnarray}
g({\bf k},\omega)&=&g^{(0)}({\bf k},\omega)+g^{(0)}({\bf k},
\omega)[\Sigma^{(h)}_{1}({\bf k},\omega)g({\bf k},\omega)
\nonumber \\
&-&\Sigma^{(h)}_{2}(-{\bf k},-\omega)\Im^{\dagger}({\bf k},
\omega)], \\
\Im^{\dagger}({\bf k},\omega)&=&g^{(0)}(-{\bf k},-\omega)
[\Sigma^{(h)}_{1}(-{\bf k},-\omega)\Im^{\dagger}(-{\bf k},-\omega)
\nonumber\\
&+&\Sigma^{(h)}_{2}(-{\bf k},-\omega)g({\bf k},\omega)],
\end{eqnarray}
\end{mathletters}
where the self-energy functions $\Sigma^{(h)}_{1}({\bf k},\omega)
=\Sigma^{(h)}_{1L}({\bf k},\omega)+\sigma_{x}\Sigma^{(h)}_{1T}
({\bf k},\omega)$ and $\Sigma^{(h)}_{2}({\bf k},\omega)=
\Sigma^{(h)}_{2L}({\bf k},\omega)+\sigma_{x}\Sigma^{(h)}_{2T}
({\bf k},\omega)$, with the longitudinal and transverse parts are
evaluated as,
\begin{mathletters}
\begin{eqnarray}
&\Sigma^{(h)}_{1L}&({\bf k},i\omega_{n})={1\over N^{2}}
\sum_{\bf p, p'}[(\gamma^{2}_{\bf p+p'+k}+t^{2})\nonumber\\
&\times&{1\over\beta}\sum_{ip_{m}}g_{L}({\bf p+k},ip_{m}
+i\omega_{n})\Pi_{LL}({\bf p},{\bf p'},ip_{m}) \nonumber \\
&+&2t\gamma_{\bf p+p'+k}{1\over\beta}\sum_{ip_{m}}g_{T}({\bf
p+k},ip_{m}+i\omega_{n})\nonumber\\
&\times&\Pi_{TL}({\bf p},{\bf p'},ip_{m})],\\
&\Sigma^{(h)}_{1T}&({\bf k},i\omega_{n})= {1\over N^{2}}
\sum_{\bf p, p'}[(\gamma^{2}_{\bf p+p'+k}+t^{2})\nonumber\\
&\times&{1\over\beta} \sum_{ip_{m}}g_{T}({\bf p+k},
ip_{m}+i\omega_{n})\Pi_{TT}({\bf p},
{\bf p'},ip_{m})\nonumber \\
&+&2t\gamma_{\bf p+p'+k}{1\over\beta}\sum_{ip_{m}}g_{L}({\bf
p+k},ip_{m}+i\omega_{n})\nonumber\\
&\times&\Pi_{LT}({\bf p},{\bf p'},ip_{m})], \\
&\Sigma^{(h)}_{2L}&({\bf k},i\omega_{n})={1\over N^{2}}\sum_{\bf
p,p'}[(\gamma^{2}_{\bf p+p'+k}+t^{2})\nonumber\\
&\times&{1\over \beta}\sum_{ip_{m}} \Im_{L}(-{\bf p-k},
-ip_{m}-i\omega_{n})\Pi_{LL}({\bf p},{\bf p'},
ip_{m})\nonumber \\
&+&2t\gamma_{\bf p+p'+k}{1\over\beta}\sum_{ip_{m}}\Im_{T}(-{\bf
p-k},-ip_{m}-i\omega_{n})\nonumber\\
&\times& \Pi_{TL}({\bf p},{\bf p'},ip_{m})],\\
&\Sigma^{(h)}_{2T}&({\bf k},i\omega_{n})={1\over N^{2}}\sum_{\bf
p,p'}[(\gamma^{2}_{\bf p+p'+k}+t^{2})\nonumber\\
&\times&{1\over \beta}\sum_{ip_{m}} \Im_{T}(-{\bf p-k},
-ip_{m}-i\omega_{n})\Pi_{TT}({\bf p},{\bf p'},
ip_{m})\nonumber \\
&+&2t\gamma_{\bf p+p'+k}{1\over \beta}\sum_{ip_{m}}\Im_{L}(-{\bf
p-k},-ip_{m}-i\omega_{n})\nonumber\\
&\times& \Pi_{LT}({\bf p},{\bf p'}, ip_{m})],
\end{eqnarray}
\end{mathletters}
where $\gamma_{\bf p+p'+k}=2t{\rm cos}(p_{x}+p'_{x}+k_{x})$, and
the spin bubbles $\Pi_{\eta,\eta'}({\bf p},{\bf p'},ip_{m})=
(1/\beta)\sum_{ip'_{m}}D^{(0)}_{\eta}({\bf p'},ip'_{m})
D^{(0)}_{\eta'}({\bf p'+p},ip'_{m}+ip_{m})$, with $\eta=L,T$ and
$\eta'=L,T$. The self-energy function $\Sigma^{(h)}_{2}({\bf k},
\omega)$ contains the pairing force and dressed holon gap
function, then it is called as the effective dressed holon gap
function, while the self-energy function $\Sigma^{(h)}_{1}({\bf k}
,\omega)$ renormalizes the MF dressed holon excitation spectrum
\cite{eliashberg}, and therefore it describes the quasiparticle
coherence. Moreover, $\Sigma^{(h)}_{2}({\bf k},\omega)$ is an even
function of $\omega$, while $\Sigma^{(h)}_{1}({\bf k},\omega)$ is
not. In this case, it is convenient to break
$\Sigma^{(h)}_{1}({\bf k},\omega)$ up into its symmetric and
antisymmetric parts as, $\Sigma^{(h)}_{1}({\bf k},\omega)=
\Sigma^{(he)}_{1}({\bf k},\omega)+\omega\Sigma^{(ho)}_{1}({\bf k},
\omega)$, then both $\Sigma^{(he)}_{1}({\bf k},\omega)$ and
$\Sigma^{(ho)}_{1}({\bf k},\omega)$ are even functions of
$\omega$. Now we define the quasiparticle coherent weights as
$Z^{(1)-1}_{FA}({\bf k},\omega)=Z^{-1}_{F1}({\bf k},\omega)-
Z^{-1}_{F2}({\bf k},\omega)$ and $Z^{(2)-1}_{FA}({\bf k},\omega)
=Z^{-1}_{F1}({\bf k},\omega)+Z^{-1}_{F2}({\bf k},\omega)$, with
$Z^{-1}_{F1}({\bf k},\omega)=1-\Sigma^{(ho)}_{1L}({\bf k},\omega)$
and $Z^{-1}_{F2}({\bf k},\omega)=\Sigma^{(ho)}_{1T}({\bf k},
\omega)$. As in the case of the planar cuprate superconductor
\cite{feng3}, the retarded function Re$\Sigma^{(he)}_{1}({\bf k},
\omega)$ is a constant, independent of $({\bf k},\omega)$, and
just renormalizes the chemical potential, therefore it can be
dropped. Since we only discuss the low-energy behavior of doped
two-leg ladder cuprates, then the effective dressed holon pair gap
functions and quasiparticle coherent weights can be discussed in
the static limit, i.e., $\bar{\Delta}_{h}({\bf k})=
\Sigma^{(h)}_{2}({\bf k},\omega)\mid_{\omega=0}=
\bar{\Delta}_{hL}({\bf k})+\sigma_{x}\bar{\Delta}_{hT}({\bf k})$,
$Z^{(1)-1}_{FA}({\bf k})=Z^{-1}_{F1}({\bf k})-Z^{-1}_{F2}({\bf k})
$ and $Z^{(2)-1}_{FA} ({\bf k})=Z^{-1}_{F1}({\bf k})+
Z^{-1}_{F2}({\bf k})$, with $Z^{-1}_{F1}({\bf k})=1-
\Sigma^{(ho)}_{1L}({\bf k},\omega)\mid_{\omega=0}$ and
$Z^{-1}_{F2}({\bf k})=\Sigma^{(ho)}_{1T} ({\bf k},\omega)
\mid_{\omega=0}$, then the longitudinal and transverse parts of
the dressed holon normal and anomalous Green's functions in Eq.
(8) can be written explicitly as,
\begin{mathletters}
\begin{eqnarray}
g_{L}({\bf k},\omega)&=&{1\over 2}\sum_{\nu=1,2}Z^{(\nu)}_{FA}
({\bf k})\left ({U^{2}_{\nu{\bf k}}\over\omega-E_{\nu{\bf k}}}+
{V^{2}_{\nu{\bf k}}\over \omega+E_{\nu{\bf k}}}\right), \\
g_{T}({\bf k},\omega)&=&{1\over 2}\sum_{\nu=1,2}(-1)^{\nu+1}
Z^{(\nu)}_{FA }({\bf k})\left ({U^{2}_{\nu{\bf k}}\over\omega-
E_{\nu{\bf k}}}\right .\nonumber\\
&+&\left . {V^{2}_{\nu{\bf k}}\over\omega+E_{\nu{\bf k}}}
\right ), \\
\Im^{\dagger}_{L}({\bf k},\omega)&=&-{1\over 2}\sum_{\nu=1,2}
Z^{(\nu)}_{FA }({\bf k}){\bar{\Delta}_{hz}^{(\nu)}({\bf k})\over
2E_{\nu{\bf k}}}\left ({1\over \omega-E_{\nu{\bf k}}}\right .
\nonumber\\
&-&\left . {1\over\omega+ E_{\nu{\bf k}}}\right ), \\
\Im^{\dagger}_{T}({\bf k},\omega)&=&-{1\over 2}\sum_{\nu=1,2}
(-1)^{\nu+1}Z^{(\nu)}_{FA }({\bf k}){\bar{\Delta}_{hz}^{(\nu)}
({\bf k})\over 2E_{\nu{\bf k}}} \left ({1\over\omega-E_{\nu{\bf
k}}}\right . \nonumber\\
&-&\left .{1\over \omega+E_{\nu{\bf k}}}\right ),
\end{eqnarray}
\end{mathletters}
where the dressed holon quasiparticle coherence factors
$U^{2}_{\nu{\bf k}}=[1+\bar{\xi}_{\nu{\bf k}}/E_{\nu{\bf k}}]/2$
and $V^{2}_{\nu{\bf k}}=[1-\bar{\xi}_{\nu{\bf k}}/E_{\nu{\bf k}}]
/2$, the renormalized dressed holon excitation spectrum
${\bar\xi}_{\nu{\bf k}}=Z^{(\nu)}_{FA}({\bf k})\xi_{\nu{\bf k}}$,
the renormalized dressed holon pair gap function
$\bar{\Delta}_{hz}^{(\nu)}({\bf k})=Z^{(\nu)}_{FA}({\bf k})[\bar{
\Delta}_{hL}({\bf k})+(-1)^{\nu+1}\bar{\Delta}_{hT}({\bf k})]$,
and the dressed holon quasiparticle dispersion $E_{\nu{\bf k}}=
\sqrt{[{\bar\xi}_{\nu{\bf k}}]^{2}+\mid\bar{\Delta}_{hz}^{(\nu)}
({\bf k})\mid^{2}}$. Although $Z^{(1)}_{FA}({\bf k})$ and
$Z^{(2)}_{FA}({\bf k})$ still are a function of ${\bf k}$, the
wave vector dependence may be unimportant, since everything
happens at the electron Fermi surface \cite{eliashberg}. In this
case, we need to estimate a special wave vector ${\bf k_{0}}$ that
guarantees $Z^{(\nu)}_{FA }=Z^{(\nu)}_{FA}({\bf k_{0}})$ near the
electron Fermi surface. Following the discussions in the case of
the planar cuprate superconductors \cite{feng3}, this special wave
vector can be obtained as ${\bf k_{0}}={\bf k_{A}}-{\bf k_{F}}$,
with ${\bf k_{A}}=\pi$ and ${\bf k_{F}}\approx(1-x)\pi/2$, then we
only need to calculate $Z_{FA}^{(\nu)}=Z_{FA}^{(\nu)}({\bf k_{0}})
$. On the other hand, many authors have shown that
superconductivity in doped two-leg ladder cuprates possesses a
modified d-wave symmetry \cite{sigrist,riera}, and the gap
function in this modified d-wave symmetry can be expressed as,
$\bar{\Delta}^{(1)}_{hz}({\bf k})=\bar{\Delta}_{hzL}({\bf k})+
\bar{\Delta}_{hzT}({\bf k})=2\bar{\Delta}_{hzL}{\rm cos}k_{x}
+\bar{\Delta}_{hzT}$ and $\bar{\Delta}^{(2)}_{hz}({\bf k})=
\bar{\Delta}_{hzL}({\bf k})-\bar{\Delta}_{hzT}({\bf k})=
2\bar{\Delta}_{hzL}{\rm cos}k_{x}-\bar{\Delta}_{hzT}$ for the
antibonding and bonding cases, respectively. In this case, the
dressed holon effective gap parameters and quasiparticle coherent
weights in Eq. (9) satisfy following four equations,
\begin{mathletters}
\begin{eqnarray}
\bar{\Delta}_{hL}&=&-{1\over 32N^{3}}\sum_{{\bf k,q,p}}
\sum_{\nu,\nu',\nu''}{\rm cos}(k_{x}-p_{x}+q_{x})\nonumber\\
&\times&\Lambda^{(1)}_{\nu\nu'\nu''}({\bf k,q,p}), \\
\bar{\Delta}_{hT}&=&-{1\over 32N^{3}}\sum_{{\bf k,q,p}}
\sum_{\nu,\nu',\nu''}(-1)^{\nu+\nu'+\nu''+1}\nonumber\\
&\times&\Lambda^{(1)}_{\nu\nu'\nu''}({\bf k,q,p}), \\
{1\over Z_{F1}}&=&1+{1\over 32N^{2}}\sum_{{\bf q,p}}
\sum_{\nu,\nu',\nu''}\Lambda^{(2)}_{\nu\nu'\nu''}({\bf q,p}),\\
{1\over Z_{F2}}&=&-{1\over 32N^{2}}\sum_{{\bf q,p}}
\sum_{\nu,\nu',\nu''}(-1)^{\nu+\nu'+\nu''+1}\nonumber\\
&\times&\Lambda^{(2)}_{\nu\nu'\nu''}({\bf q,p}),
\end{eqnarray}
\end{mathletters}
where the kernel functions $\Lambda^{(1)}_{\nu\nu'\nu''}({\bf k,
q,p})$ and $\Lambda^{(2)}_{\nu\nu'\nu''}({\bf q,p})$ are evaluated
as,
\begin{mathletters}
\begin{eqnarray}
\Lambda^{(1)}_{\nu\nu'\nu''}({\bf k,q,p})&=&
{Z^{(\nu'')2}_{FA}B_{\nu'{\bf p}} B_{\nu{\bf q}}\over
\omega_{\nu'{\bf p}} \omega_{\nu{\bf q}}}[\gamma_{{\bf k+q}}+
t(-1)^{\nu+\nu''}]^{2}\nonumber\\
&\times&[2\bar{\Delta}_{hL}{\rm cos}k_{x}+
\bar{\Delta}_{hT}(-1)^{\nu''+1}]\nonumber \\
&\times&\left ({F^{(1)}_{\nu\nu'\nu''}({\bf q,p})+
F^{(2)}_{\nu\nu'\nu''}({\bf k,q,p})\over [\omega_{\nu'{\bf p}}
-\omega_{\nu{\bf q}}]^2-E^{2}_{\nu''{\bf k}}}\right . \nonumber\\
&+&\left . {F^{(3)}_{\nu\nu'\nu''}({\bf q,p})+
F^{(4)}_{\nu\nu'\nu''}({\bf k, q,p})\over [\omega_{\nu'{\bf p}}+
\omega_{\nu{\bf q}}]^2-E^{2}_{\nu''{\bf k}}} \right ),\\
\Lambda^{(2)}_{\nu\nu'\nu''}({\bf q,p})&=&{Z^{(\nu'')}_{FA}
B_{\nu'{\bf p}}B_{\nu{\bf q}}\over \omega_{\nu'{\bf p}}
\omega_{\nu{\bf q}}}[\gamma_{\bf p+k_{0}}+t(-1)^{\nu+\nu''}]^{2}
\nonumber\\
&\times&\left({H^{(1)}_{\nu\nu'\nu''}({\bf q,p})\over
[\omega_{\nu'{\bf p} }-\omega_{\nu{\bf q}}+E_{\nu''{\bf
p-q+k_{0}}}]^{2}} \right .
\nonumber \\
&+&{H^{(2)}_{\nu\nu'\nu''}({\bf q,p})\over [\omega_{\nu'{\bf p}}
-\omega_{\nu{\bf q}}-E_{\nu''{\bf p-q+k_{0}}}]^{2}}\nonumber\\
&+&{H^{(3)}_{\nu\nu'\nu''}({\bf q,p})\over [\omega_{\nu'{\bf p}}
+\omega_{\nu{\bf q}}+E_{\nu''{\bf p-q+k_{0}}}]^{2}}
\nonumber \\
&+&\left . {H^{(4)}_{\nu\nu'\nu''}({\bf q,p})\over [\omega_{\nu'
{\bf p}} +\omega_{\nu{\bf q}}-E_{\nu''{\bf p-q+k_{0}}}]^{2}}
\right ),
\end{eqnarray}
\end{mathletters}
respectively, where $F^{(1)}_{\nu\nu'\nu''}({\bf q, p})= n_{B}
(\omega_{\nu{\bf q}})+n_{B}(\omega_{\nu'{\bf p}})+2n_{B}
(\omega_{\nu{\bf q}})n_{B}(\omega_{\nu'{\bf p}})$,
$F^{(2)}_{\nu\nu'\nu''}({\bf k,q,p})=[2n_{F}(E_{\nu''{\bf k}})
-1][\omega_{\nu'{\bf p}}-\omega_{\nu{\bf q}}][n_{B}
(\omega_{\nu{\bf q}})-n_{B}(\omega_{\nu'{\bf p}})]/E_{\nu''{\bf
k}}$, $F^{(3)}_{\nu\nu'\nu''}({\bf q,p})=1+n_{B}(\omega_{\nu{\bf
q}})+n_{B}(\omega_{\nu'{\bf p}})+2n_{B}(\omega_{\nu{\bf q}})
n_{B}(\omega_{\nu'{\bf p}})$, $F^{(4)}_{\nu\nu'\nu''}({\bf k,q,p})
=[2n_{F}(E_{\nu''{\bf k}}) -1][\omega_{\nu'{\bf p}}+
\omega_{\nu{\bf q}}][1+n_{B}(\omega_{\nu{\bf q}})+
n_{B}(\omega_{\nu'{\bf p}})]/E_{\nu''{\bf k}}$,
$H^{(1)}_{\nu\nu'\nu''}({\bf q,p})=n_{F}(E_{\nu''{\bf p-q+k_{0}}})
[n_{B}(\omega_{\nu'{\bf p}})-n_{B}(\omega_{\nu{\bf q}})]+
n_{B}(\omega_{\nu{\bf q}})[1+n_{B}(\omega_{\nu'{\bf p}})]$,
$H^{(2)}_{\nu\nu'\nu''}({\bf q,p})=n_{F}(E_{\nu''{\bf p-q+k_{0}}})
[n_{B}(\omega_{\nu{\bf q}})-n_{B}(\omega_{\nu'{\bf p}})]+
n_{B}(\omega_{\nu'{\bf p}})[1+n_{B}(\omega_{\nu{\bf q}})]$,
$H^{(3)}_{\nu\nu'\nu''}({\bf q, p})=[1-n_{F}(E_{\nu''{\bf
p-q+k_{0}}})][1+n_{B}(\omega_{\nu{\bf q}}) +n_{B}(\omega_{\nu'{\bf
p}})]+n_{B}(\omega_{\nu{\bf q}})n_{B}(\omega_{\nu'{\bf p}})$, and
$H^{(4)}_{\nu\nu'\nu''}({\bf q,p})=n_{F}(E_{\nu''{\bf p-q+k_{0}}})
[1+n_{B}(\omega_{\nu{\bf q}})+n_{B}(\omega_{\nu'{\bf p}})]+
n_{B}(\omega_{\nu{\bf q}})n_{B}(\omega_{\nu'{\bf p}})$. These four
equations (11) must be solved together with other self-consistent
equations,
\begin{mathletters}
\begin{eqnarray}
\phi_{\parallel}&=&{1\over 4N}\sum_{\nu,{\bf k}}Z^{(\nu)}_{FA}
{\rm cos}k_{x}\left (1-{{\bar\xi}_{\nu{\bf k}}\over E_{\nu{\bf
k}}}{\rm th}[{1\over 2}\beta E_{\nu{\bf k}}] \right ), \\
\phi_{\perp}&=&{1\over 4N}\sum_{\nu,{\bf k}}(-1)^{\nu+1}
Z^{(\nu)}_{FA}\left (1-{{\bar\xi}_{\nu{\bf k}}\over E_{\nu{\bf
k}}}{\rm th}[{1\over 2}\beta E_{\nu{\bf k}}]\right ), \\
\delta&=&{1\over 4N}\sum_{\nu,{\bf k}} Z^{(\nu)}_{FA} \left (
1-{{\bar\xi}_{\nu{\bf k}}\over E_{\nu{\bf k}}}{\rm th}[{1\over
2}\beta E_{\nu{\bf k}}]\right ), \\
\chi_{\parallel}&=&{1\over 4N}\sum_{\nu,{\bf k}}{\rm cos}k_{x}
{B_{\nu{\bf k}}\over\omega_{\nu{\bf k}}}{\rm coth}
[{1\over 2}\beta\omega_{\nu{\bf k}}], \\
C_{\parallel}&=&{1\over 4N}\sum_{\nu,{\bf k}}{\rm cos}^{2}k_{x}
{B_{\nu{\bf k}}\over\omega_{\nu{\bf k}}}{\rm coth}
[{1\over 2}\beta\omega_{\nu{\bf k}}], \\
1\over 2&=&{1\over 4N}\sum_{\nu,{\bf k}}{B_{\nu{\bf k}}\over
\omega_{\nu{\bf k}}}{\rm coth}[{1\over 2}\beta
\omega_{\nu{\bf k}}], \\
\chi^{z}_{\parallel}&=&{1\over 4N}\sum_{\nu,{\bf k}}{\rm cos}k_{x}
{B_{z\nu{\bf k}}\over\omega_{z\nu{\bf k}}}{\rm coth}
[{1\over 2}\beta\omega_{z\nu{\bf k}}], \\
C^{z}_{\parallel}&=&{1\over 4N}\sum_{\nu,{\bf k}}{\rm cos}^{2}
k_{x}{B_{z\nu{\bf k}}\over\omega_{z\nu{\bf k}}}{\rm coth}
[{1\over 2}\beta\omega_{z\nu{\bf k}}], \\
\chi_{\perp}&=&{1\over 4N}\sum_{\nu,{\bf k}}(-1)^{\nu+1}
{B_{\nu{\bf k}}\over\omega_{\nu{\bf k}}}{\rm coth}[{1\over
2}\beta\omega_{\nu{\bf k}}], \\
C_{\perp}&=&{1\over 4N}\sum_{\nu,{\bf k}}(-1)^{\nu+1}{\rm cos}
k_{x}{B_{\nu{\bf k}}\over\omega_{\nu{\bf k}}}
{\rm coth}[{1\over 2}\beta\omega_{\nu{\bf k}}], \\
\chi^{z}_{\perp}&=&{1\over 4N}\sum_{\nu,{\bf k}}(-1)^{\nu+1}
{B_{z\nu{\bf k}}\over\omega_{z\nu{\bf k}}}{\rm coth}[{1\over
2}\beta\omega_{z\nu{\bf k}}], \\
C^{z}_{\perp}&=&{1\over 4N}\sum_{\nu,{\bf k}}(-1)^{\nu+1}{\rm cos}
k_{x}{B_{z\nu{\bf k}}\over\omega_{z\nu{\bf k}}}{\rm coth}[{1\over
2}\beta\omega_{z\nu{\bf k}}],
\end{eqnarray}
\end{mathletters}
then all the above dressed holon effective gap parameters,
quasiparticle coherent weights, dressed holon particle-hole order
parameters, decoupling parameter $\alpha$, spin correlation
functions, and chemical potential $\mu$ are determined by the
self-consistent calculation \cite{qin}. With the helps of the
above discussions, we now obtain the longitudinal and transverse
parts of the dressed holon pair order parameter in Eq. (4) in
terms of the dressed holon anomalous Green's functions (10c) and
(10d) as,
\begin{mathletters}
\begin{eqnarray}
\Delta_{hL}&=&{1\over 2N}\sum_{\nu,{\bf k}}Z^{(\nu)}_{FA}{\rm cos}
k_{x}{\bar{\Delta}_{hz}^{(\nu)}({\bf k})\over E_{\nu{\bf k}}}
{\rm th}[{1\over 2}\beta E_{\nu{\bf k}}], \\
\Delta_{hT}&=&{1\over 2N}\sum_{\nu,{\bf k}}(-)^{\nu+1}
Z^{(\nu)}_{FA}{\bar{\Delta}_{hz}^{(\nu)}({\bf k})\over E_{\nu{\bf
k}}}{\rm th}[{1\over 2}\beta E_{\nu{\bf k}}] .
\end{eqnarray}
\end{mathletters}
As in the case of the planar cuprate superconductors
\cite{feng2,feng3}, this dressed holon pairing state originating
from the kinetic energy terms by exchanging the spin excitations
in doped two-leg ladder cuprates also leads to form the electron
Cooper pairing state \cite{feng2}. For discussions of
superconductivity in doped two-leg ladder cuprates, we need to
calculate the electron anomalous Green's function
$\Gamma^{\dagger}({\bf k},\omega)=\Gamma^{\dagger}_{L}({\bf k},
\omega)+\sigma_{x}\Gamma^{\dagger}_{T}({\bf k},\omega)$, with the
longitudinal and transverse parts are defined as,
$\Gamma^{\dagger}_{L} (i-j,t-t')=\langle\langle
C^{\dagger}_{ia\uparrow}(t);C^{\dagger}_{ja\downarrow}(t')\rangle
\rangle$ and $\Gamma^{\dagger}_{T}(i-j,t-t')=\langle\langle
C^{\dagger}_{ia\uparrow}(t);C^{\dagger}_{ja'\downarrow}(t')\rangle
\rangle(a'\neq a)$. These longitudinal and transverse parts of the
electron anomalous Green's function are the convolutions of the
corresponding longitudinal and transverse parts of the dressed
holon anomalous Green's function and spin Green's function in the
CSS fermion-spin theory, and reflect the charge-spin recombination
\cite{anderson1}. In terms of the MF spin Green's functions in
Eqs. (6c) and (6d) and dressed holon anomalous Green's functions
(10c) and (10d), we obtain the longitudinal and transverse parts
of the electron anomalous Green's function, then the longitudinal
and transverse parts of the SC gap function $\Delta({\bf
k})=\Delta_{L}({\bf k})+ \sigma_{x}\Delta_{T}({\bf k}) $ are
evaluated as,
\begin{mathletters}
\begin{eqnarray}
\Delta_{L}(\bf k)&=&-{1\over 16}{1\over N}\sum_{{\bf p},\nu,\nu'}
Z^{(\nu')}_{FA}{\bar{\Delta}_{hz}^{(\nu')}({\bf p-k})\over
E_{\nu'{\bf p-k}}}{B_{\nu{\bf p}}\over\omega_{\nu{\bf
p}}}\nonumber \\
&\times&{\rm th} [{1\over 2}\beta E_{\nu'{\bf p-k}}]{\rm coth}
[{1\over 2}\beta\omega_{\nu{\bf p}}], \\
\Delta_{T}(\bf k)&=&-{1\over 16}{1\over N}\sum_{{\bf p},\nu,\nu'}
(-1)^{\nu+\nu'}Z^{(\nu')}_{FA}{\bar{\Delta}_{hz}^{(\nu')}({\bf
p-k})\over E_{\nu'{\bf p-k}}}{B_{\nu{\bf p}}\over \omega_{\nu{\bf
p}}}\nonumber \\
&\times&{\rm th}[{1\over 2}\beta E_{\nu'{\bf p-k}}]{\rm coth}
[{1\over 2}\beta\omega_{\nu{\bf p}}],
\end{eqnarray}
\end{mathletters}
which shows that the symmetry of the electron Cooper pair in doped
two-leg ladder cuprates is essentially determined by the symmetry
of the dressed holon pairs. In this case, the SC gap function is
written as ${\Delta}^{(1)}({\bf k})={\Delta}_{L}({\bf k})+
{\Delta}_{T}({\bf k})=2{\Delta}_{L}{\rm cos}k_{x}+{\Delta}_{T}$
and ${\Delta}^{(2)}({\bf k})={\Delta}_{L}({\bf k})-
{\Delta}_{T}({\bf k})=2{\Delta}_{L}{\rm cos}k_{x}-{\Delta}_{T}$
for the antibonding and bonding cases, respectively, then the
longitudinal and transverse parts of the SC gap parameter are
evaluated in terms of Eqs. (15) and (14) as $\Delta_{L}=-
\chi_{\parallel}\Delta_{hL}$ and $\Delta_{T}=-\chi_{\perp}
\Delta_{hT}$. In Fig. 1, we plot the longitudinal (solid line) and
transverse (dashed line) parts of the dressed holon pairing (a)
and SC (b) gap parameters as a function of the doping
concentration $\delta$ for parameter $t/J=2.5$ at temperature
$T=0.0001J$. Our result shows that both longitudinal and
transverse parts have almost the same amplitude, and the
longitudinal (transverse) part of the dressed holon pairing
parameter has a similar doping dependent behavior of the
longitudinal (transverse) part of the SC gap parameter. In
particular, the value of the longitudinal part of the SC gap
parameter $\Delta_{L}$ increases with increasing doping in the
underdoped regime, and reaches a maximum in the optimal doping
$x_{{\rm opt}}\approx 0.07$, then decreases in the overdoped
regime.

\begin{figure}[prb]
\epsfxsize=3.7in\centerline{\epsffile{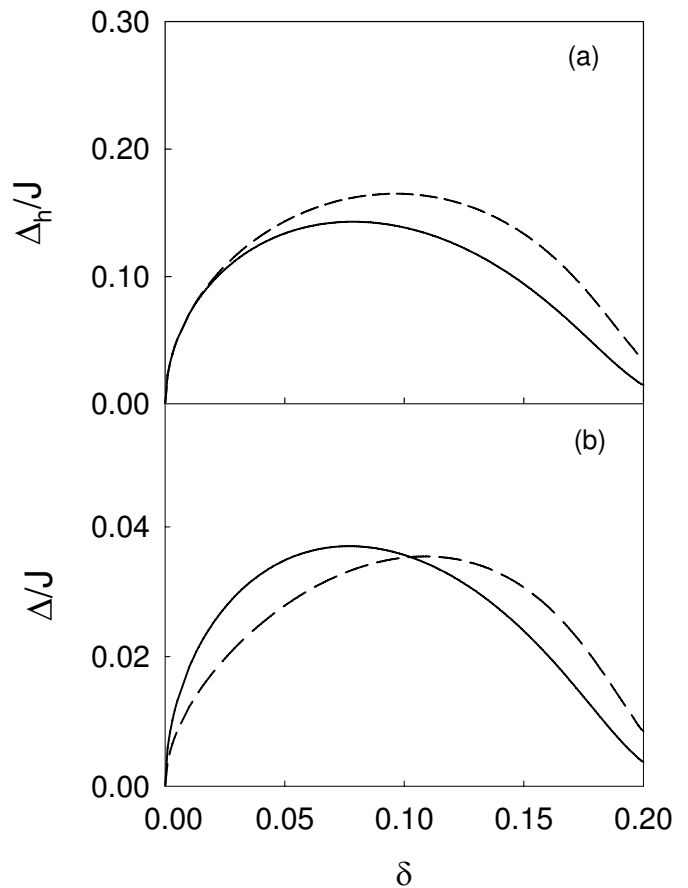}} \caption{The
longitudinal (solid line) and transverse (dashed line) dressed
holon pairing (a) and SC (b) gap parameters as a function of the
doping concentration for $t/J=2.5$ with $T=0.0001J$. }
\end{figure}

\begin{figure}[prb]
\epsfxsize=3.7in\centerline{\epsffile{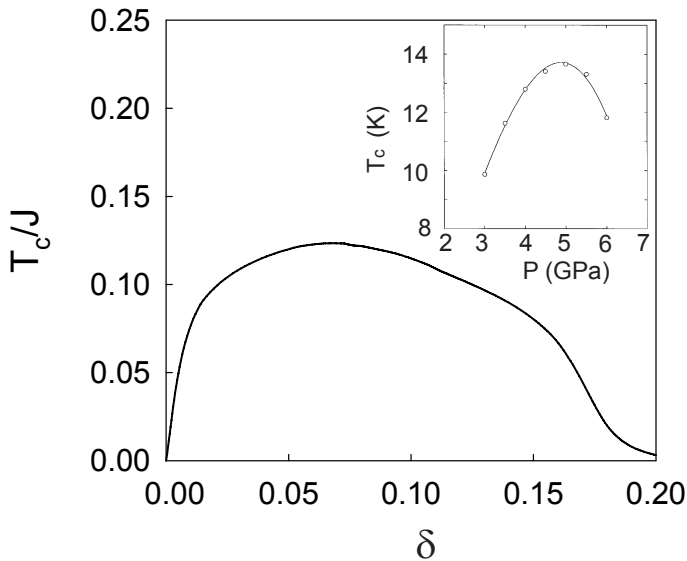}} \caption{ The SC
transition temperature as a function of the doping concentration
for $t/J=2.5$. Inset: the experimental result taken from Ref. [7].
}
\end{figure}

Our result in Eq. (15) also shows that the SC transition
temperature $T_{c}$ occurring in the case of the SC gap parameter
$\Delta=0$ is identical to the dressed holon pair transition
temperature occurring in the case of the dressed holon pairing gap
parameter $\Delta_{h}=0$. In correspondence with the SC gap
parameter, the SC transition temperature $T_{c}$ as a function of
the hole doping concentration $\delta$ for $t/J=2.5$ is plotted in
Fig. 2. For comparison, the experimental result of the SC
transition temperature $T_{c}$ in the doped two-leg ladder cuprate
Sr$_{0.4}$Ca$_{13.6}$Cu$_{24}$O$_{41}$ as a function of pressure
\cite{isobe} is also shown in Fig. 2 (inset). Experimentally, it
has been shown that the main effect of pressure in doped two-leg
ladder cuprates is to reduce the distance between the ladders and
chains, which leads to the doped hole redistribution between
chains and ladders \cite{uehara,nagata,isobe}. However, the
structure at ambient pressure with clearly defined ladders and
chains remains stable under the pressure needed to induce the SC
regime \cite{isobe}, and the spin background in this pressure
induced SC phase does not drastically alter its spin gap
properties \cite{dagotto2}. On the other hand, when Ca is doped
upon the original Sr-based Ca-undoped phase, the interatomic
distance ladder-chain was found to be reduced by Ca substitution,
leading to a redistribution of holes originally present only on
the chains \cite{ohta,kato}. These experimental results show that
an increase of the pressure may corresponding to an increase in
the number of charge carriers on the ladders
\cite{isobe,ohta,kato}. In other words, the doping dependence of
the SC transition temperature should be similar to the pressure
dependence of the SC transition temperature. In this sense, our
present result of the doping dependence of the SC transition
temperature is qualitatively consistent with the experimental
result \cite{isobe}. The maximal SC transition temperature T$_{c}$
occurs around the optimal doping concentration $x_{{\rm opt}}
\approx 0.07$, and then decreases in both underdoped and overdoped
regimes. In particular, this domed shape of the doping dependence
of the SC transition temperature is same as that of the doping
dependence of the longitudinal part of the SC gap parameter, which
shows that superconductivity is mainly produced by the development
of the pairing correlation along legs, and is consistent with the
one-dimensional charge dynamics under high pressure
\cite{isobe,dagotto2}. Furthermore, T$_{c}$ in the underdoped
regime is proportional to the hole doping concentration $\delta$,
and therefore T$_{c}$ in the underdoped regime is set by the hole
doping concentration, which reflects that the density of the
dressed holons directly determines the superfluid density in the
underdoped regime.

The essential physics of superconductivity in the present doped
two-leg ladder cuprate superconductors is the same as that in the
planar cuprate superconductors \cite{feng2,feng3}, i.e., the
SC-order in doped two-leg ladder cuprate superconductors is
controlled by both gap function and quasiparticle coherence, which
is reflected explicitly in the self-consistent equations (11). The
dressed holons (then electrons) interact by exchanging the spin
excitations and that this interaction is attractive. This
attractive interaction leads to form the dressed holon pairs (then
electron Cooper pairs). The parent compound of doped two-leg
ladder cuprates is a Mott insulator, when holes are doped into
this insulator, there is a gain in the kinetic energy per hole
proportional to $t$ due to hopping, but at the same time, the spin
correlation is destroyed, costing an energy of approximately $J$
per site, therefore the doped holes into the Mott insulator can be
considered as a competition between the kinetic energy ($\delta
t$) and magnetic energy ($J$), and the magnetic energy decreases
with increasing doping. In the underdoped and optimally doped
regimes, the magnetic energy is rather large, then the dressed
holon (then electron) attractive interaction by exchanging the
spin excitations is also rather strong to form the dressed holon
pairs (then electron Cooper pairs) for the most dressed holons
(then electrons), therefore the number of the dressed holon pairs
(then electron Cooper pairs) and SC transition temperature are
proportional to the hole doping concentration. However, in the
overdoped regime, the magnetic energy is relatively small, then
the dressed holon (then electron) attractive interaction by
exchanging the spin excitations is also relatively weak, in this
case, not all dressed holons (then electrons) can be bounden as
dressed holon pairs (then electron Cooper pairs) by this weak
attractive interaction, and therefore the number of the dressed
holon pairs (then electron Cooper pairs) and SC transition
temperature decrease with increasing doping.

In summary, we have discussed superconductivity with the modified
d-wave symmetry in doped two-leg ladder cuprates based on the
kinetic energy driven SC mechanism. It is shown that the
spin-liquid ground-state at the half-filling evolves into the SC
ground-state upon doping. In analogy to the doping dependence of
the SC transition temperature in the planar cuprate
superconductors, the SC transition temperature in doped two-leg
ladder cuprates increases with increasing doping in the underdoped
regime, and reaches a maximum in the optimal doping, then
decreases in the overdoped regime.

When this work was finished we became aware of the discovery of
the analogous quasi-one dimensional cuprate superconductor
\cite{sasaki} Pr$_{2}$Ba$_{4}$Cu$_{7}$O$_{15-\delta}$ at ambient
pressure. It has been shown \cite{sasaki} that in addition to the
CuO$_{2}$ planes, the cuprate superconductor
Pr$_{2}$Ba$_{4}$Cu$_{7}$O$_{15-\delta}$ contains two CuO chains.
In particular, these two single chains bound together like the
two-leg ladder in two-leg ladder cuprates. This cuprate
superconductor Pr$_{2}$Ba$_{4}$Cu$_{7}$O$_{15-\delta}$ is called a
double chain SC system, since the quasi-one dimensional double
chain turns into the SC-state at ambient pressure, while the
planes remain insulating even below the SC transition temperature.
Although the Cu ions in the double chain do not form the two-leg
ladder structure but a zigzag chain, the quasi-one dimensional
nature of the double chain in the cuprate superconductor
Pr$_{2}$Ba$_{4}$Cu$_{7}$O$_{15-\delta}$ is the same as that of the
two-leg ladder in two-leg ladder cuprates. This experimental
measurement \cite{sasaki} on
Pr$_{2}$Ba$_{4}$Cu$_{7}$O$_{15-\delta}$ provides an evidence that
the doped quasi-one dimensional cuprates can become SC at ambient
pressure when the doped charge carriers distribute properly along
the chains. These and related issue is under investigation now.

\acknowledgments The authors would like to thank Dr. Ying Liang,
Dr. Tianxing Ma, Dr. Bin Liu, and Dr. Huaiming Guo for the helpful
discussions. This work was supported by the National Natural
Science Foundation of China under Grant Nos. 10547104, 10125415,
and 90403005, and the funds from the Ministry of Science and
Technology of China under Grant No. 2006CB601002.

\end{document}